\documentclass[twocolumn,prl]{revtex4}

\usepackage[pdftex]{graphicx}
\usepackage{dcolumn} % Align table columns on decimal point
\usepackage{subfigure}
\usepackage{times}

\def\simlt{\mathrel{\lower .3ex \rlap{$\sim$}\raise .5ex \hbox{$<$}}}
\def\simgt{\mathrel{\lower .3ex \rlap{$\sim$}\raise .5ex \hbox{$>$}}}

\def\and{ {\wedge} }

\begin{document}

\title{The computational complexity of Kauffman nets and the 
P versus NP problem}

\author{S. N. Coppersmith}

\affiliation{Department of Physics, University of Wisconsin-Madison, 1150 
University Avenue, Madison, WI 53706}

\date{\today}

\begin{abstract}
Complexity theory as practiced by physicists and computational complexity 
theory as practiced by computer scientists both characterize how difficult
it is to solve complex problems.
Here it is shown that the parameters of a specific model
can be adjusted so that the problem of finding its global energy minimum
is extremely
sensitive to small changes in the problem statement.
This result has implications not only for studies of the physics of
random systems but may also lead to new strategies for resolving
the well-known P versus NP question in computational complexity theory.
\end{abstract}

\pacs{75.10.Nr, 89.20.Ff, 02.60.Pn}
%05.45.-a Nonlinear dynamics and nonlinear dynamical systems
%05.10.Cc Renormalization group methods
%05.90.+m other topics in statistical physics, thermodynamics, and
%nonlinear dynamical systems
%89.20.Ff Computer science and technology
%02.60.Pn Numerical optimization
%75.10.Nr Spin-glass and other random models

\maketitle
Computational complexity theory addresses the question of
how fast the resources required to solve a given problem grow 
with the size of the input needed to specify the problem.\cite{papadimitriou94}
The close relationships between the physics of random systems and
computational complexity theory have been explored for nearly two
decades.~\cite{fu86,mezard87}

Whether or not P, the complexity class of problems that can be solved 
in a time that grows polynomially with the size of the problem 
specification (``polynomial time''), and NP, the class of problems 
for which a solution can be verified in polynomial time, are distinct
is a central unanswered question in computational 
complexity theory.\cite{claymath}
The class of NP-complete problems are equivalent in
that being able to solve any one of them in polynomial time implies 
that any problem in NP can be solved in polynomial time.\cite{cook71,levin86,garey79}
An intuitive picture believed to be appropriate for NP-complete 
problems is that the presence of conflicting constraints, or 
``frustration,''\cite{toulouse77} causes each problem to have an ``energy landscape'' 
with many local minima, and finding the global minimum is difficult 
because typical algorithms must explore an extremely large number 
of local minima to find the global one.\cite{mezard87}

Here it is shown that the energy landscapes
of NP-complete problems have important differences in
their {\em local} properties.
Specifically, it is demonstrated that 
the parameters of a particular problem in NP can 
be adjusted so that the solution is extremely sensitive to small 
changes in the problem, and it is argued that exploiting this sensitivity 
may lead to new strategies for resolving the P versus NP question.

We consider here a specific system called a Kauffman 
net, or \textit{N}-\textit{K} model.\cite{kauffman69}
It has \textit{N} Boolean elements
$\sigma _{i} $, \textit{i}=1, 2,..., \textit{N}, and
the value of the $i^{th}$ element  at time \textit{t}+1 is determined by the 
value of its \textit{K} inputs 
$j_{1} (i),$ $j_{2} (i),$ $...,$ $j_{K} (i)$ at time \textit{t} via 
\begin{equation}
\sigma _{i} (t+1)=f_{i} (\sigma _{j_{1} (i)} (t), \sigma _{j_{2} (i)}
(t),\ldots,\sigma _{j_{K} (i)} (t)),
\label{eq:model_def}
\end{equation}
where each $f_i$ is a randomly chosen Boolean function that depends 
on \textit{K} arguments.
The \textit{K} inputs for each element and the 
Boolean functions $f_i$ are all chosen randomly before beginning and 
then fixed throughout the computation.
Specifying a Kauffman net requires time and space polynomial 
in \textit{N} so long as \textit{K} grows no faster than log(\textit{N)}: 
this follows because specifying the \textit{K} inputs for each of 
the \textit{N} elements takes a number of bits proportional to \textit{NK} log(\textit{N}), 
and specifying the $f_i$ takes $N2^K$ bits (one per 
output for each of the $2^K$ possible inputs for each element).
For the predecessor problem, 
a natural choice for the energy of a given configuration \{\ensuremath{\tau}\} 
is the number of bits in the successor configuration
\{\ensuremath{\sigma}\}=\textit{f}(\{\ensuremath{\tau}\}) 
that differ from the corresponding bit in the target configuration.\cite{baskaran87}

Here we consider the problem of determining
whether a given configuration of a Kauffman net
$\{ {\sigma} \}=\{\sigma_1,\sigma_2,\ldots,\sigma_N\} $
has a predecessor.
When $K=N$, although 
specifying the model 
requires space that grows exponentially with \textit{N}, 
one can still ask how many evaluations of the Boolean functions
are required 
to determine whether such a predecessor exists.
A candidate solution can be verified with a single evaluation of
each Kauffman net function, but
because in this case each configuration is a truly random function of 
its predecessor\cite{derrida86,derrida87},
%finding 
%the solution if it is not supplied in advance requires exponentially 
%many evaluations.~\cite{bennett81}
%This result 
%%is intuitively plausible because 
%follows because
%the successor configurations are chosen randomly and independently, 
%so that 
the only way to determine whether a predecessor exists is 
to check all of the exponentially many candidates.~\cite{bennett81}

%As discussed above,
When $K=A\log_2N$ with \textit{A} a constant, the problem of determining whether 
a given configuration has a predecessor is in NP.
%, for the problem 
%can be specified in a space and a solution can be verified 
%in a time that grows polynomially with \textit{N}. 
The problem is potentially hard because
one must determine whether many different, potentially 
conflicting constraints can be satisfied simultaneously. Goldreich\cite{goldreich00}
has speculated that finding a predecessor requires 
time exponential in $N$, and the question of whether a given configuration of a 
given Kauffman net has a predecessor can easily be rewritten 
as an instance of satisfiability, a classic NP-complete problem.\cite{garey79}
The instances of satisfiability that correspond to randomly chosen 
Kauffman nets are expected to require exponential time
for any \textit{K}\ensuremath{\geq}3.\cite{mezard02}

For any $K>2$, local 
search algorithms that work by decreasing the number of wrong 
bits by changing a small number of bits in the current ``guess'' 
of the predecessor typically find only a local minimum and not 
the global one.
A new ingredient that enters for $K=A\log_2N$ is that finding the global 
minimum is hard even if one starts off with a configuration whose 
successor has only one bit in error. In contrast, if one has 
found a configuration whose successor differs from the target 
by one bit for a random problem instance with \textit{K}=3, the error 
can be corrected with very high probability via a small number 
of single bit flips.
In addition, when \textit{K}\ensuremath{=}\textit{\ensuremath{A}}log$_{2}$\textit{N} 
with \textit{A} large enough, if the target configuration with the 
single bit flip has a predecessor, the number of bits in the 
predecessor configuration that must be changed grows as a power 
of \textit{N}, while when K=3 a change in a single bit in the target 
results in a change of only O(1) bits of the predecessor.

The sensitivity of the predecessor configuration to single-bit 
changes in the target constrains the computational strategies 
that could solve the predecessor problem in polynomial time. 
First, the algorithm must yield the exact answer, since the local 
search algorithm cannot correct even single-bit errors. Second, 
the algorithm must explicitly depend on the specification of 
every bit of every input function as well as every bit of the 
target configuration. This is because if one realization of the 
functions yields the target output, then a second function realization 
that differs from the first by a single bit change could yield 
a configuration that differs from the target by one bit, and 
the arguments given above then demonstrate that the predecessor 
for the second function realization, if it exists, has a large 
number of bits that are different than for the first function 
realization.

The extreme sensitivity of the predecessor configuration to small 
changes in the target suggests that a strategy for providing 
a lower bound on the difficulty of solving the predecessor problem 
could be to investigate the complexity of the polynomial describing 
the predecessor function. For example, one could define a function 
$g_{i}(\sigma_{1},\sigma_{2},\ldots,\sigma_{N})$ to be zero 
if the configuration $\{\sigma\}$ has a single predecessor $\{\tau\}$ 
with \ensuremath{\tau}$_{i}$=0, one if \{\ensuremath{\sigma}\} has a single predecessor 
\{\ensuremath{\tau}\} with \ensuremath{\tau}$_{i}$=1, two if the configuration \{\ensuremath{\sigma}\} 
has no predecessor, and three if \{\ensuremath{\sigma}\} has more than 
one predecessor. The degree of this polynomial is bounded below 
by the number of its zeros, and since changes of a single bit 
in the target change many bits of the predecessor, the number 
of zeros of this polynomial grows approximately as fast as the 
number of target configurations that have a single predecessor. 
Moreover, the sensitive dependence of the polynomial's coefficients 
on the choice of the Kauffman net functions constrain the possible 
compact algorithms for computing them. Implementing this strategy 
is nontrivial, since one must prove that the number 
of target configurations with a single predecessor grows faster 
than polynomially with $N$ as well as that the large number of 
different possible polynomials precludes efficient computation 
of the coefficients. Nonetheless, considering random networks 
with $K\propto\log N$ is useful in this context because the sensitivity 
of the predecessor to single bit changes in the successor enables 
one to relate the number of sign changes of the function to the 
number of configurations with a single predecessor.

To demonstrate the increasing sensitivity of  the 
solution of the predecessor problem
to single-bit changes in
the target as $K$ is increased, one assumes that a
one is given a configuration 
\{\ensuremath{\tau}\} such that \textit{f}(\{\ensuremath{\tau}\})\ensuremath{\equiv}\{\ensuremath{\eta}\} 
differs from the target configuration $\{ \sigma \}$ by exactly one bit.
One then attempts to find the state \{\ensuremath{\nu}\} such that \textit{f}(\{\ensuremath{\nu}\})=\{\ensuremath{\sigma}\} as follows:
(1) For each of the \textit{K} inputs of the wrong element \textit{i}, find 
the configuration that results when a given input is flipped, 
and 
(2) flip the input of \textit{i} that minimizes the number of wrong 
elements in the output. 
This ``downhill'' algorithm succeeds if single-element changes 
of \{\ensuremath{\tau}\} yield no errors in the output instead of one 
error. 

This algorithm is characterized here
using methods similar to those in~\cite{bastolla96,bastolla98}.
Starting with a configuration that yields a successor that differs 
from the target by one bit, the probability that flipping 
one given input of the wrong bit causes the output result to 
have \textit{k} errors is 
\begin{equation} 
P_{K} (k)=2^{-K} \frac{K!}{k!(K-k)!} .
\label{eq:P_K}
\end{equation}
Eq.~(\ref{eq:P_K}) follows because a given input affects \textit{K} elements, 
and the output for each is correct with probability 1/2.

Now one gets to pick the input that yields the fewest incorrect 
outputs. Defining $Q_K(j)$ as the probability that the best output configuration 
differs from the target in \textit{j} bits, one finds 
\begin{widetext}
\begin{eqnarray}
Q_{K} (0) &=& 1-(1-P_{K} (0))^{K}\nonumber\\
Q_{K} (j) &=& \left( 1-\sum_{i=0}^{j-1}P_{K} (i)\right) ^{K} -\left( 1-\sum_{i=0}^{j}P_{K} (i)\right) ^{K} \qquad 1 \le j\le K-1 \nonumber \\
Q_{K} (K) &=& \left( 1-\sum\limits_{j=0}^{K-1}P_{K} (j) \right) ^{K} ~.
\label{eq:Q_K}
\end{eqnarray}
\end{widetext}

Eqs.~(\ref{eq:Q_K}) follow because if the smallest number 
of errors yielded by this process is \textit{i}, no trials can yield 
a number of errors less than \textit{i}, at least one trial must yield \textit{i} 
errors, and 
$\left( 1-\int\limits_{j=0}^{i}P_{K} (j) \right) ^{K} $ is the probability that 
only more than \textit{i} errors are obtained.

This procedure fixes the error if the resulting configuration has no
wrong bits.
If the number of wrong bits is one, 
then it is likely that a different element \ensuremath{\ell} is now 
wrong, and one can repeat the procedure and perhaps find a correct 
solution in the next iteration. If the number of wrong bits is 
more than one, then the process has yielded a configuration that 
is farther from a solution.

One way to estimate whether the process is getting closer to 
or farther from a solution is to calculate after one step of 
the procedure the expectation value of the number of wrong bits, 
$\left\langle N_{error} (K)\right\rangle $, 
\begin{equation}
\left\langle N_{error} (K)\right\rangle =\sum\limits_{j=1}^{K}KQ_{K} (j)
~.
\label{eq:N_error}
\end{equation}

Now $\left\langle N_{error} (K)\right\rangle $
 increases monotonically with \textit{K}, and
$\left\langle N_{error} (K=3)\right\rangle =0.797$
 and 
$\left\langle N_{error} (K=4)\right\rangle =1.005$, so the 
number of wrong bits is reduced by the procedure when \textit{K}=3 
but not when \textit{K}=4.
Thus, it appears that the procedure corrects 
the one-bit error with high probability for some values of \textit{K} 
strictly greater than 3, but not when \textit{K} is larger than 4.

The energy landscape of the model becomes
increasingly ``rugged'' as \textit{K} is increased.~\cite{kauffman93},
and the probability that the downhill algorithm fixes the error
decreases monotonically as $K$ increases.
When $K=A\log_2 N$, the number of wrong bits 
after one step is strongly peaked at 
$K/2=(A/2)\log _{2} N$,
and the probability that the operation yields either zero errors or
one error $\propto N^{-A}$, and
the probability of being able to correct \textit{any} 
one-bit error vanishes as
$N\rightarrow \infty $.

Now it is shown that if the target configuration that is changed by one bit 
has a predecessor, for large enough $A$
the new predecessor configuration differs 
from the original one by a number of bits that grows at
least as fast as
\textit{N}$^{\mathit{x}}$, with \textit{x} strictly greater than zero.
This follows because
(1) changing a single input bit is extremely likely 
to change many output bits, and (2)
if one chooses \textit{M} random numbers out of \textit{N} possibilities, 
when \textit{M} is much smaller than 
$\sqrt{N}$, the probability of duplication is exponentially small in the 
ratio $M/\sqrt{N}$.\cite{diaconis89}

Here, we consider
Kauffman nets in which each element has 
exactly \textit{K} inputs and \textit{K} outputs, a restriction that does 
not affect any of the results but simplifies the analysis.
One 
begins with a configuration \{\ensuremath{\tau}\} such that f(\{\ensuremath{\tau}\}) 
is the target configuration \{\ensuremath{\sigma}\}, and now considers 
a new target configuration \{\ensuremath{\sigma}'\} that is identical to 
\{\ensuremath{\sigma}\} except for a single bit flip. One then asks how 
many bits one must flip in the configuration \{\ensuremath{\tau}\} to 
obtain a configuration
\{\ensuremath{\tau}'\} such that f(\{\ensuremath{\tau}'\})=\{\ensuremath{\sigma}'\}. 
To see that \{\ensuremath{\tau}'\} differs from \{\ensuremath{\tau}\} by many bits, 
consider all configurations that differ from the original predecessor 
\{\ensuremath{\tau}\} by \textit{S} bits, where S \texttt{<} \textit{N}$^{x}$, with \textit{x}\texttt{<}1/2. 
The number of different ways that these \textit{S} bits can be chosen 
is 
$N!/\left[ S!\left( N-S\right) !\right] $, which can be written 
$\left( Ne/S\right) ^{S} +O\left( S^{2} /N\right) $
 when \textit{N} and \textit{S} are large and \textit{S\texttt{<}\texttt{<}}
$\sqrt{N} $. Next we define \textit{Q} to be the number of elements with at least 
one input that has been changed. When \textit{S\texttt{<}\texttt{<}}
$\sqrt{N} /K$
 and \textit{K} is large, \textit{Q\texttt{>}SK/}2. This is because the number 
of affected outputs would be \textit{SK} if none were duplicated, 
and while one can obtain individual duplications of an output 
by choosing inputs that have a common output, the probability 
of obtaining additional duplications when the connections are 
chosen randomly is negligible so long as \textit{SK}\texttt{<}\texttt{<}
$\sqrt{N} $.
Now we define \textit{R} to be the number of elements whose output 
value has changed. When \textit{Q} is large, given a realization with \textit{Q} 
outputs that have at least one perturbed input, the probability 
distribution governing the number of output bits that are different 
from the original target is a normal distribution with mean \textit{Q/}2 
and standard deviation $\sqrt{Q/2}$.\cite{feller68}
Therefore, the probability that \textit{R\texttt{<}Q/}4 for such 
a realization must be less than 
$(Q/4)\exp \left[ \left( Q/4-Q/2\right) ^{2} /2Q\right] $
$=\left( Q/4\right) \exp \left( -Q^{2} /32\right) $.

So now finally we show that when $A$ is large enough the probability that a choice of 
the \textit{S} inputs exists for which \textit{R\texttt{<}SK/}8 is smaller than \textit{1/N} 
as \textit{N}\textit{\ensuremath{\rightarrow}}\textit{\ensuremath{\infty}}.
This follows because the probability 
that \textit{Q\texttt{>}SK/}2 is essentially unity and that the probability 
that \textit{R\texttt{<}QK/}4 is very small. Specifically, the conclusion 
follows if
\begin{equation}
\left( \frac{Ne}{S} \right) ^{S} Q\exp \left[ -Q/32\right] <1~.
\label{eq:condition1}
\end{equation}

For \textit{Q\texttt{>}SK}/2 and \textit{K=A}log$_{2}$\textit{N}, one has
\begin{eqnarray}
\left( \frac{Ne}{S} \right) ^{S} Q e^{  -\frac{Q}{32}} &<& \left(
\frac{Ne}{S} \right) ^{S} \left( \frac{SK}{2} \right) e^{ \left[
-\frac{SK}{64}\right]} \nonumber \\
  &=& \left( \frac{Ne}{S} \right) ^{S} \left( \frac{SK}{2}
\right) N^{-\frac{SA}{64\ln 2} } \nonumber\\
&=& \left( \frac{Ne}{S} \right) ^{S} \left( \frac{SK}{2} \right) \left(
\frac{1}{N^{\frac{A}{64\ln 2}} } \right) ^{S} ~.½
\label{eq:condition2}
\end{eqnarray}

The expression (\ref{eq:condition2}) becomes arbitrarily small as
\textit{N}\textit{\ensuremath{\rightarrow}}\textit{\ensuremath{\infty}} 
for any \textit{S\texttt{<}N}$^{\mathit{x}}$ with \textit{x\texttt{<}}1/2 when \textit{A}/64 ln \textit{2 
\texttt{>}}1.

Thus, we have shown that no configuration that differs from the 
unperturbed predecessor by a number of bits that is smaller than \textit{S}\textit{\ensuremath{\propto}}\textit{N}$^{\mathit{x}}$ 
with \textit{x\texttt{<}}1/2 can be the predecessor of a new target configuration 
that differs from the original target by \textit{O(}1\textit{)} bits.

To summarize, we have shown that the problem of finding a predecessor 
configuration of a random Kauffman net in which each element 
has a number of connections that grows logarithmically with the 
number of elements is extremely sensitive to single-bit changes 
in the target configuration. We are optimistic that this sensitivity 
to small changes can be exploited to develop new strategies to 
attack the difficult question of whether the complexity P and 
NP are distinct. 

The author gratefully acknowledges financial support from grants 
NSF-DMR 0209630 and NSF-EMT 0523680, and useful conversations 
with Eric Bach, Robert Joynt, and Dieter von Melkebeek. The hospitality 
of the Aspen Center for Physics, where some of this work was 
done, is greatly appreciated.

\end{document}